\documentclass[12pt]{article}
\usepackage{epsfig}
\usepackage{amsfonts}
\usepackage{amsopn}
\usepackage{amsmath}

\title{
\vskip -50 pt
\begin{flushright}
\normalsize\rm NORDITA-2011-5
\end{flushright}
\vskip 20 pt
On the mass term of the Dirac equation}

\author{
Maciej Trzetrzelewski \thanks{e-mail: maciej.trzetrzelewski@gmail.com} \\ \\
NORDITA,  \\
Roslagstullsbacken 23, 106 91 Stockholm, \\
Sweden \\ \\
Department of Mathematics,\\
Royal Institute of Technology, \\
KTH, 100 44 Stockholm, \\
Sweden \\
}

\begin{document}
\date{}
\maketitle

\abstract{We consider the generalization of the Dirac equation where the mass
term is an arbitrary matrix $M$. A general form of $M$, consistent with the mass shell constraint, is derived and proven to be equivalent to the original Dirac equation.}

\section{Introduction}
The original way \cite{Dirac} in which Dirac obtained relativistic
equation for fermions seems to leave certain ambiguities related to
the choice of the mass term. This led some authors \cite{leitner} to
discuss the possibility of generalizing the term by considering
certain mass matrices $M$ instead of the usual matrix $m
\bold{1}$. We would like to point out that consistency conditions
actually imply that $M$ must be given by $M=m e^{(i \alpha-\beta)
\gamma^5}$ with $\alpha \in [0,2\pi]$ and $\beta \in \mathbb{R}$, of
which the cases $\beta=0$ and $\alpha=0$  were discussed in
\cite{leitner}. This mass term can be obtained from the Dirac
equation by an appropriate
 change of the phases and the norms of the Weyl spinors.

The presented derivation is done entirely in one reference system i.e. we do not use explicitly Poincar\'{e} invariance
(nowhere in the derivation we use the fact that the Dirac equation should have the same form in all reference frames).   \\

\section{General mass term}
Consider a non hermitian $x^{\mu}$ dependent matrix $M$ and assume
that the corresponding Dirac equations $D_M\psi=0$,
$D_M=-i\gamma^{\mu}\partial_{\mu}+M$ holds. There exist a simple
argument following from Poincar\'{e} invariance that $M$ has to be of the
form $M=a+b\gamma^5$, $a,b \in \mathbb{C}$: first $M$ has to be
independent of $x^{\mu}$ due to the translation invariance of the
equation $D_M\psi=0$, second multiplying the equation $D_M\psi=0$
from the l.h.s. and the r.h.s. by $U$ and $U^{-1}$ respectively,
where $U$ is the Lorentz transformation, and using he fact that in
the new reference frame the equation should have the same form, one
concludes that $[M,\gamma^{\mu\nu}]=0$ (where
$\gamma^{\mu\nu}=\frac{1}{2}[\gamma^{\mu},\gamma^{\nu}]$) which has
the general solution $M=a+b\gamma^5$. To find expression for $a$ and
$b$ one has to use the mass-shell constraint.

We would like to point out that the explicit use of Poincar\'{e} covariance of equation  $D_M\psi=0$ is not necessary to prove that $M=a+b\gamma^5$. To see this 
let us note that for an arbitrary operator
$\mathcal{D}$ the consistency conditions $\mathcal{D}D_M\psi=0$ have
to be satisfied. Due to the mass shell constraint
$p_{\mu}p^{\mu}=m^2$, $p_{\mu}=-i\partial_{\mu}$, useful conditions
will come from operators $\mathcal{D}$ which involve the
$i\gamma^{\mu}\partial_{\mu}$ operator. Let us consider
\begin{equation} \label{1}
0=D_{-M}D_M\psi=(m^2-M^2-i\gamma^{\mu}\partial_{\mu}M)\psi-i[\gamma^{\mu},M]\partial_{\mu}\psi
\end{equation}
(we use the conventions $\eta_{\mu\nu}=diag(1,-1,-1,-1)$,
$\{\gamma^{\mu},\gamma^{\nu}\}=2\eta^{\mu\nu}\bold{1}$, $\gamma^5=+i\gamma^0\gamma^1\gamma^2\gamma^3$). One can also consider other equations e.g.
\begin{equation} \label{other}
0=D_{-M^{\dagger}}D_M\psi=D_{M}D_M\psi=D_{M^{\dagger}}D_M\psi=D_{\bold{0}}D_M\psi
\end{equation}
however as it turns out they do not give new constraints.

If $M$ is equal to $m\bold{1}$, equation (\ref{1}) is
trivially satisfied (equations in (\ref{other}) are either trivial or give the Dirac equation $D_m
\psi=0$). For general $M$ we obtain some nontrivial, first order,
differential equations for $\psi$. These equations must reduce to
the Dirac equation $D_M\psi=0$ - otherwise we would obtain an
independent equation for fermions. Concentrating on Eqn. (\ref{1})
we conclude that
\begin{equation} \label{6}
[\gamma^{\mu},M]=A\gamma^{\mu},
\end{equation}
\begin{equation} \label{7}
m^2-M^2-i\gamma^{\mu}\partial_{\mu}M=AM
\end{equation}
for some matrix $A$.

In order to solve (\ref{6}) and (\ref{7}) it is useful to multiply
Eqn. (\ref{6}) from the r.h.s. by $\gamma^{\mu}$ (no sum) which implies the following equations
\[
\gamma^iM\gamma^i-\gamma^jM\gamma^j=0, \ \ \ \ \gamma^0M\gamma^0+\gamma^jM\gamma^j=0 \ \ \ \ 1 \le i<j\le 3
\]
or simply $[M,\gamma^{\mu\nu}]=0$ hence $M=a(x)+b(x)\gamma^5$.
To obtain the condition for $a$ and $b$ we use (\ref{6}) to find that $A=-2b(x)\gamma^5$ hence (\ref{7}) gives
\begin{equation} \label{9}
m^2=a(x)^2-b(x)^2+\gamma^{\mu}\partial_{\mu}\left(a(x)+b(x)\gamma^5\right).
\end{equation}
The r.h.s. in (\ref{9}) should be proportional to the unit matrix
hence  $\partial_{\mu}a=\partial_{\mu}b=0$. Therefore the general
solution of (\ref{6}) and (\ref{7}) and hence of the constraint
(\ref{1}) is given by
\[
M=a+b\gamma^5, \ \ \ \ a,b \in \mathbb{C},
\]
\begin{equation} \label{10}
m^2=a^2-b^2
\end{equation}
which also solves other constraint (\ref{other}). Therefore we
obtained the desired condition as a consequence of consistency
equations. While the mass-shell constraint used in this
derivation is Poincar\'{e} invariant, let us note that we did not
used explicitly its symmetries.

Using the parametrization for the complex circle in (\ref{10})
\[
a=m(\cos \alpha \cosh \beta - i \sin \alpha \sinh \beta),
\]
\[
b=m i (\sin \alpha \cosh \beta + i \cos \alpha \sinh \beta)
\]
with $\alpha \in [0,2\pi]$ and $\beta \in \mathbb{R}$ we can write $M$ in the compact form
\begin{equation} \label{15}
M=m e^{(i\alpha-\beta) \gamma^5}.
\end{equation}

Finally let us observe that this form of $M$ can be obtained from the Dirac equation with $M=m \bold{1}$. Noting that in the Weyl representation we have
\[
i \sigma^{\mu}\partial_{\mu} \psi_L =me^{-i\alpha+\beta} \psi_R,
\]
\[
i \bar{\sigma}^{\mu}\partial_{\mu} \psi_R =me^{i\alpha-\beta} \psi_L
\]
where $\psi_R$, $\psi_L$ are Weyl spinors and choosing
$\tilde{\psi}_L=e^{\frac{i\alpha -\beta}{2}}\psi_L, \
\tilde{\psi}_R=e^{-\frac{i\alpha -\beta}{2}}\psi_R$  (which could be
interpreted as the chiral transformation with the complex angle) the
Dirac equation for Weyl spinors $\tilde{\psi}_R$, $\tilde{\psi}_L$
transforms into the standard form with
$M=m$. \\

A particularly interesting  $M$ not belonging to (\ref{15}) is $M=m\gamma_0$.
The condition (\ref{1}) implies that   $i \gamma^j\partial_j \psi =0$ and hence
$(i\partial_0 -m)\psi=0$ which clearly is not Lorentz invariant.
There are no negative energy solutions for this
choice i.e. the plane-wave ansatz $\psi=u e^{-i kx}$ for positive
energy solutions and $\psi=v e^{i kx}$ for negative energy
solutions, implies $k_i=0$, $k_0=m$ for four basis spinors
$[u_s]_t=\delta_{st}$, $s,t=1,2,3,4$ and no solutions for the $v$
spinor.

\vspace{12pt} \noindent{\bf Acknowledgments}

I thank D. Lundholm for comments.


\begin{thebibliography}{7}


\bibitem{Dirac}
P. A. M. Dirac, \emph{The Quantum Theory of the Electron}, Proc. of the Royal Society, Series A, Vol. 117, No. 778 (1928), pp. 610-624.

\bibitem{leitner}
D. Leitner, G. Szamosi, \emph{Pseudoscalar Mass and Its Relationship to Conventinal Scalar Mass of the Relativistic Dirac Theory of the Electron}, Lettere al Nuovo Cimento, vol. 5, No. 12, (1972), 814-816.



\end{thebibliography}
\end{document}